\documentclass[twocolumn]{aastex63}
\usepackage{rotating}
\usepackage{graphicx}
\usepackage{amssymb}
\usepackage{amsmath}
\usepackage{hyperref}

\def\cxo{{\sl CXO\ }}


\def\gtrsim{\mathrel{\hbox{\rlap{\hbox{\lower4pt\hbox{$\sim$}}}\hbox{$>$}}}}
\def\lesssim{\mathrel{\hbox{\rlap{\hbox{\lower4pt\hbox{$\sim$}}}\hbox{$<$}}}}
\def\farcs{\hbox{$.\!\!^{\prime\prime}$}}

\begin{document}

\title{A dragon out of breath? Monitoring high-velocity outflows from the high-mass gamma-ray binary LS 2883/PSR B1259-63 during the 2017--2021 binary cycle}
\author{Jeremy Hare}
\affil{Astrophysics Science Division, NASA Goddard Space Flight Center, 8800 Greenbelt Rd, Greenbelt, MD 20771, USA}
\affiliation{Center for Research and Exploration in Space Science and Technology, NASA/GSFC, Greenbelt, Maryland 20771, USA}
\affiliation{The Catholic University of America, 620 Michigan Ave., N.E. Washington, DC 20064, USA}
\author{George G. Pavlov}
\affiliation{Department of Astronomy \& Astrophysics, Pennsylvania State University, 525 Davey Lab, University Park, PA 16802, USA}
\author{Gordon P.\ Garmire}
\affiliation{Huntingdon Institute for X-ray Astronomy, LLC, 10677 Franks Road, Huntingdon, PA 16652, USA}
\author{Oleg Kargaltsev}
\affiliation{Department of Physics, The George Washington University, 725 21st St. NW, Washington, DC 20052}
\affiliation{The George Washington Astronomy, Physics, and Statistics Institute of Sciences (APSIS)}
\email{jeremy.hare@nasa.gov}

\begin{abstract}

Observations of the high-mass gamma-ray binary LS 2883/PSR B1259--63 with the Chandra X-ray Observatory during the 2011--2014 and 2014--2017 binary cycles  have shown  
X-ray emitting 
clumps, 
presumably ejected from the binary during 
periastron passages.
These clumps traveled at 
 projected velocities of $\sim0.1 c$  and have shown evidence of being accelerated. The clumps also evolved in shape, size, and flux.
We monitored this binary with Chandra during the 2017--2021 binary cycle to search for additional X-ray emitting ejections. While we find evidence of extended emission in two of the six observations, it is unlike the clumps observed in the previous three binary cycles. More specifically, the extended emission is not well localized and no bright clump is observed moving away from the binary. It is still unclear what caused the lack of X-ray emitting clump in this orbital cycle, but it may be due to changes in the decretion disk of the Be star.

\end{abstract}

\section{Introduction}
High-mass gamma-ray binaries (HMGBs) 
consist of a massive O or B type star being orbited by 
a neutron star (NS) or a black hole.
From three HMGBs, namely LS\,2883/PSR\,B1259--63,  MT91\,213/PSR\,J2032+4127, and LS I\,+61$^{\circ}$\,303/PSR\,J0240+6113 (\citealt{1992ApJ...387L..37J,2009ApJ...705....1C,2022NatAs...6..698W}), radio pulsations have been detected, confirming a NS as the compact object. Observationally, 
such systems differ from accreting  high-mass X-ray binaries (HMXBs) hosting NSs in that their broadband emission peaks at MeV/GeV energies instead of 
hard X-ray energies as in HMXBs (see e.g., \citealt{2013A&ARv..21...64D}). In HMGBs hosting a NS, it is thought that the X-ray emission and $\gamma$-ray emission 
are due to synchrotron 
and inverse Compton (IC) radiation, respectively, emitted by leptons accelerated at the shock 
created by the colliding  pulsar and stellar winds. This suggests that the NSs in these systems are young enough to still produce strong winds, and that these winds are energetic enough to prevent material from reaching the NS surface, unlike in HMXBs. This is further supported by the observed radio pulsations from these systems, which are quenched in accreting NS HMXBs.

The 
best-studied 
HMGB is LS 2883 hosting the 48 ms pulsar PSR B1259--63 (collectively referred to as B1259 hereafter), 
and located at a distance of 2.6 kpc.
The pulsar is orbiting a 15--30 $M_{\odot}$ rapidly rotating Be star 
in a highly eccentric ($e=0.87$) orbit with a period $P_{\rm orb}=1236.7$ days
\citep{2011ApJ...732L..11N,2018MNRAS.479.4849M}. PSR B1259-63 has an energy loss rate $\dot{E}\approx8\times10^{35}$ erg cm$^{-2}$ s$^{-1}$, characteristic age
$\tau\approx330$ kyr, and 
a magnetic field $B=3.3\times10^{11}$ G \citep{1992ApJ...387L..37J}. 
Twice per orbit, it plunges through the decretion disk of the companion Be star,  which is estimated to be inclined by $\approx$35$^{\circ}$ to the 
orbital plane (see e.g., \citealt{1999ApJ...514L..39B,2005MNRAS.358.1069J,2015MNRAS.454.1358C}).

Extended X-ray emission, seen up to $\sim4''$ from B1259 near apastron with Chandra, was first reported by \cite{2011ApJ...730....2P}. During the next orbital cycle, between the periastrons of 2011 and 2014, Chandra observed the system three more times and in each of the observations, once again found an extended emission feature. 
This feature, nicknamed ``the clump", 
was moving away from the binary
along the major axis of the orbit in the periastron-to-apastron direction with a projected velocity of $v_{\perp}\approx0.08d_{2.6}$ $c$, where $d=2.6d_{2.6}$ kpc is the distance to the binary \citep{2014ApJ...784..124K,2015ApJ...806..192P}. 
Between the periastron passages of 2014 and 2017, Chandra 
observed the binary five more times.
Similar to  the previous binary cycle, high velocity (i.e., $v_{\perp}\approx0.15d_{2.6}\,c$) extended emission was found moving  
in the same direction, but this time the extended emission showed evidence of being accelerated with a projected acceleration, $a_{\perp}\sim 50$
cm s$^{-2}$ \citep{2019ApJ...882...74H}. Additionally, this extended emission changed shape and showed strong flux variability, including an epoch where it brightened by about a factor of two, before fading.

The current 
explanation for the motion, velocity, and X-ray emission from the clump is as follows. The pulsar spends the majority of 
the binary period near apastron due to 
the high eccentricity of the orbit. If, as is likely, the 
pulsar to stellar wind momentum flux ratio, $\eta\equiv \dot{E}/(\dot{M_*}v_{w} c)<1$, where $\dot{M_*}$ is the Be star's mass-loss rate to its isotropic wind, and $v_w$ is the stellar wind velocity, then the 
pulsar wind (PW) will be confined by the stellar wind.  
The PW  then carves out a channel in
the ambient medium along the periastron-to-apastron direction. When the pulsar passes through the decretion disk shortly after periastron, some material from the stellar decretion disk may get knocked into the channel carved out by the PW. In this scenario, the X-ray emission is due to synchrotron emission from particles accelerated at the interface between the clump and PW (see \citealt{2014ApJ...784..124K,2015ApJ...806..192P,2019ApJ...882...74H}, for more detailed discussions). Additionally, the interaction between the PW and clump can accelerate the clump to the observed projected velocities of $\sim10\%$ $c$ over timescales of $\approx1000$ days  \citep{2016MNRAS.456L..64B}. 

Here we report on the analysis and results of the Chandra observations of this system during the 2017 Sep 22 -- 2021 Feb 9 orbital cycle. This paper is organized as follows. In Section \ref{obs_and_dat} we present the observations and data reduction. In Section \ref{analysis} we present our analysis of the data, and the results. The discussion can be found in Section \ref{discuss}, and we end with concluding remarks in Section \ref{concc}.

\begin{table*}
\label{tab1}
\caption{Spectral fit parameters for the core and extended emission in  six \cxo ACIS imaging observations} 
\begin{center}
\renewcommand{\tabcolsep}{0.11cm}
\begin{tabular}{lcccccccccccc}
\tableline 
ObsID	&  MJD &	$\theta$\tablenotemark{a} 	&	$\Delta t$\tablenotemark{b}	&	Exp.\tablenotemark{c}	&	Cts\tablenotemark{d}	& 
$F_{\rm obs}$\tablenotemark{e}	&	$F_{\rm corr}$\tablenotemark{f}	&	$N_H$	&	$\Gamma$	&	$\mathcal{N}$\tablenotemark{g}	&
$\mathcal{A}$\tablenotemark{h} 	&	$\chi^2/$dof \\
&  &	deg	&	days	&	ks	&	&	$10^{-14}$ cgs	&	$10^{-14}$ cgs	&	$10^{21}$ cm$^{-2}$	&	&
$10^{-4}$	&	arcsec$^2$	&\\
\tableline 
21245 & 58482 & 173 &463.5 & 34.3 & 2329 & 129(4) & 139(4) & 1.6(9) & 1.33(7) & 1.7(2) & 7.1 & 63.7/77\\
21246 & 58629 & 180 & 610.6 & 37.1 & 2147 & 98(3) & 123(5) & 4.2(9) & 1.66(8) & 2.1(2) & 7.1 & 114.3/70\\
  & & & & & 38 (2\,bkg) & 1.7 & 1.8 & 1.6$^\ast$ & 1.4$^\ast$ &  & 5.1 &  \\
21247 & 58783 & 186 &  764.4 & 37.1 &  1948 & 92(3) & 122(5) & 3(1) & 
1.71(9) & 2.0(2) & 7.1 & 61.2/65 \\
   & & & & &  21 (2\,bkg) &  0.9 & 1.0 & 1.6$^\ast$ &  1.4$^\ast$ &   & 7.3  & \\
22693 & 59026 & 200 & 1008.3 & 42.6 & 4109 &  167(4) & 200(7) & 4.7(7) & 1.77(6) & 4.1(3) & 7.1 & 161.7/130 \\
22694 & 59079 & 205 & 1061.1 & 44.5 & 5954 &  235(4) & 
314(9) & 5.4(6) & 1.75(5) & 5.8(4) & 7.1 & 181.3/169 \\
   & & & & & 38 (3\,bkg) & 1.3 & 1.5 & 1.6$^\ast$ & 1.4$^\ast$ & & 10.9 & \\
comb\tablenotemark{i} & 59209 & 235 & 1190.5 & 65.0 & 10203 &  307(4) & 386(7) & 5.9(5) & 1.48(4) & 5.6(3) & 7.1 & 342.3/329 \\
\tableline 
\end{tabular} 
\end{center}
\tablecomments{ For each ObsID  the upper and the lower rows correspond to  the partially resolved core and extended emission, respectively.
The $1\sigma$ uncertainties of the last significant digits are shown in parentheses. 
Fluxes and counts are in the 0.5--8\,keV energy range. Asterisks mark fixed parameter values.}
\tablenotetext{\textnormal{a}}{True anomaly counted from periastron.}
\tablenotetext{\textnormal{b}}{Days since  latest preceded periastron. }
\tablenotetext{\textnormal{c}}{Exposure time.}
\tablenotetext{\textnormal{d}}{Total (gross) counts. The estimated numbers of background counts in the extended emission regions are shown in parentheses.}
\tablenotetext{\textnormal{e}}{Observed flux.}
\tablenotetext{\textnormal{f}}{Unabsorbed flux.}
\tablenotetext{\textnormal{g}}{Normalization in photons\,s$^{-1}$\,cm$^{-2}$\,keV$^{-1}$ at 1\,keV.}
\tablenotetext{\textnormal{h}}{Area of the extraction region.}
\tablenotetext{\textnormal{i}}{ObsIDs 23437, 24905, 24862 fit simultaneously.}
\end{table*}

\section{Observations and Data Reduction}
\label{obs_and_dat}

\begin{figure}
\includegraphics[trim={0 0 0 0},width=9.0cm]{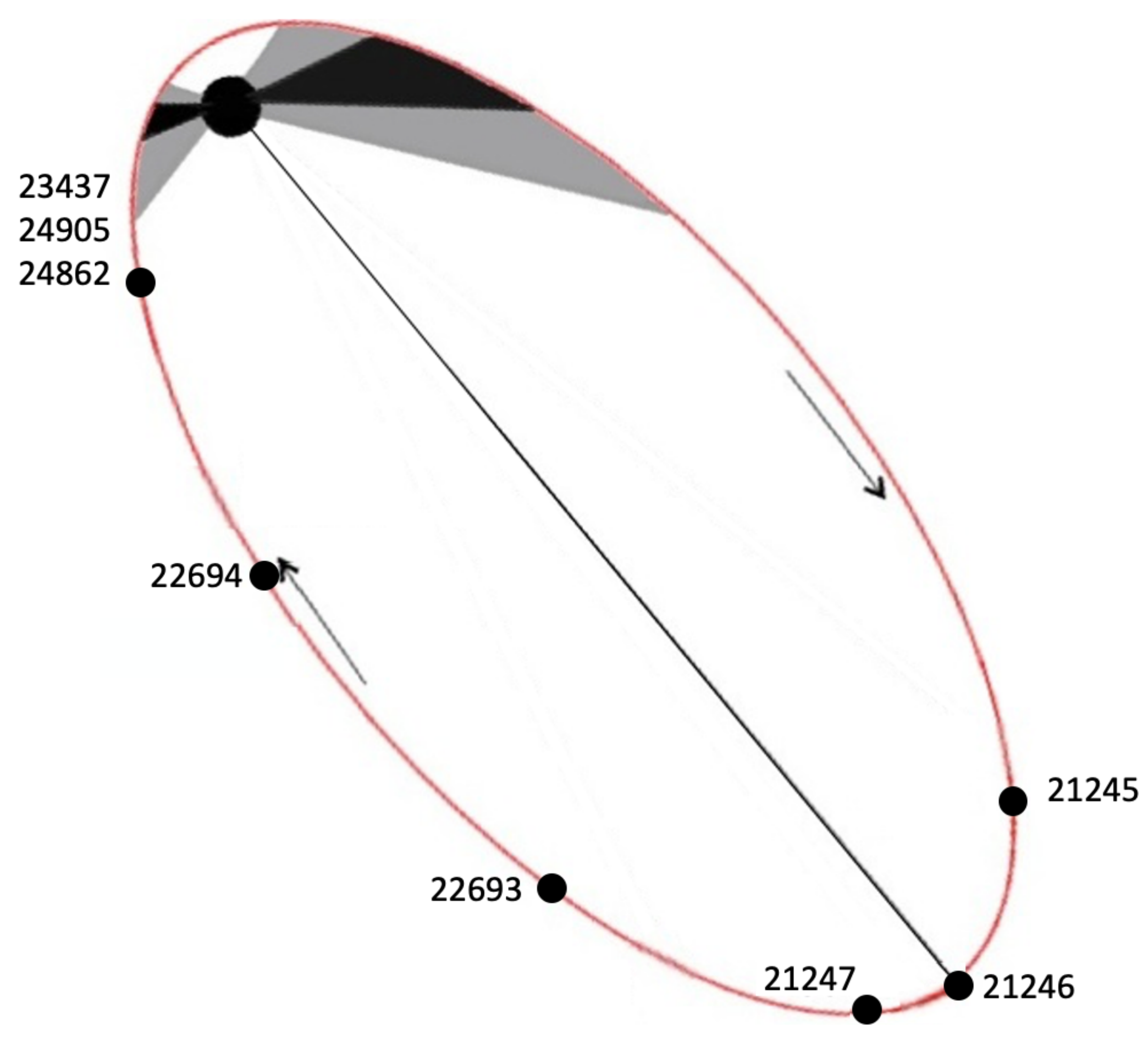}
\caption{Location of the pulsar in its orbit around the massive companion during each Chandra observation for this cycle. The gray/black
shaded regions show where in the orbit the pulsar passes
through the massive star’s decretion disk \citep{2006MNRAS.367.1201C}.
{
\label{obs_orbit}
}}
\end{figure}

During the 2017--2021 orbital cycle, 
B1259 was observed by Chandra eight times,
between 2018 December 29 and 2020 December 19.
The binary was imaged on the front-illuminated 
I3 chip of the ACIS detector in
timed exposure mode. A 1/8 sub-array was used, providing a frame time of 0.4 s, to reduce pile-up from the bright binary. The data were telemetered using the ``very faint'' format. The details 
for all eight observations are shown in Table \ref{tab1}. We note that the last three observations were taken within a three-day window. Therefore, we combine these data sets for analysis, leaving six total observing epochs. The location of the pulsar in its orbit at these epochs is shown in Figure \ref{obs_orbit}.

We used the reprocessed pipeline-produced Level 2 event files 
and filtered them to only contain photons with energies in the 0.5--8 keV range to reduce background. To extract the source spectra and detector response files, we used the Chandra Interactive Analysis ({CIAO}) version 4.14 and calibration database (CALDB) version 4.9.4. The spectra for the binary were extracted from $r=1\farcs5$ 
circular regions centered on the source. The background spectra were extracted from source-free $r\approx25''$ circular regions offset from the binary to ensure that no extended emission from the binary was used for background subtraction. All spectra were fit using Xspec version 12.13.0c \citep{1996ASPC..101...17A} from HEASoft version 6.31, with 
photo-ionization cross-sections from \cite{1996ApJ...465..487V} and the 
abundances of \cite{2000ApJ...542..914W}.

\section{Image and Spectral Analysis}
\label{analysis}

\begin{figure*}
\includegraphics[trim={0 0 0 0},width=18.0cm]{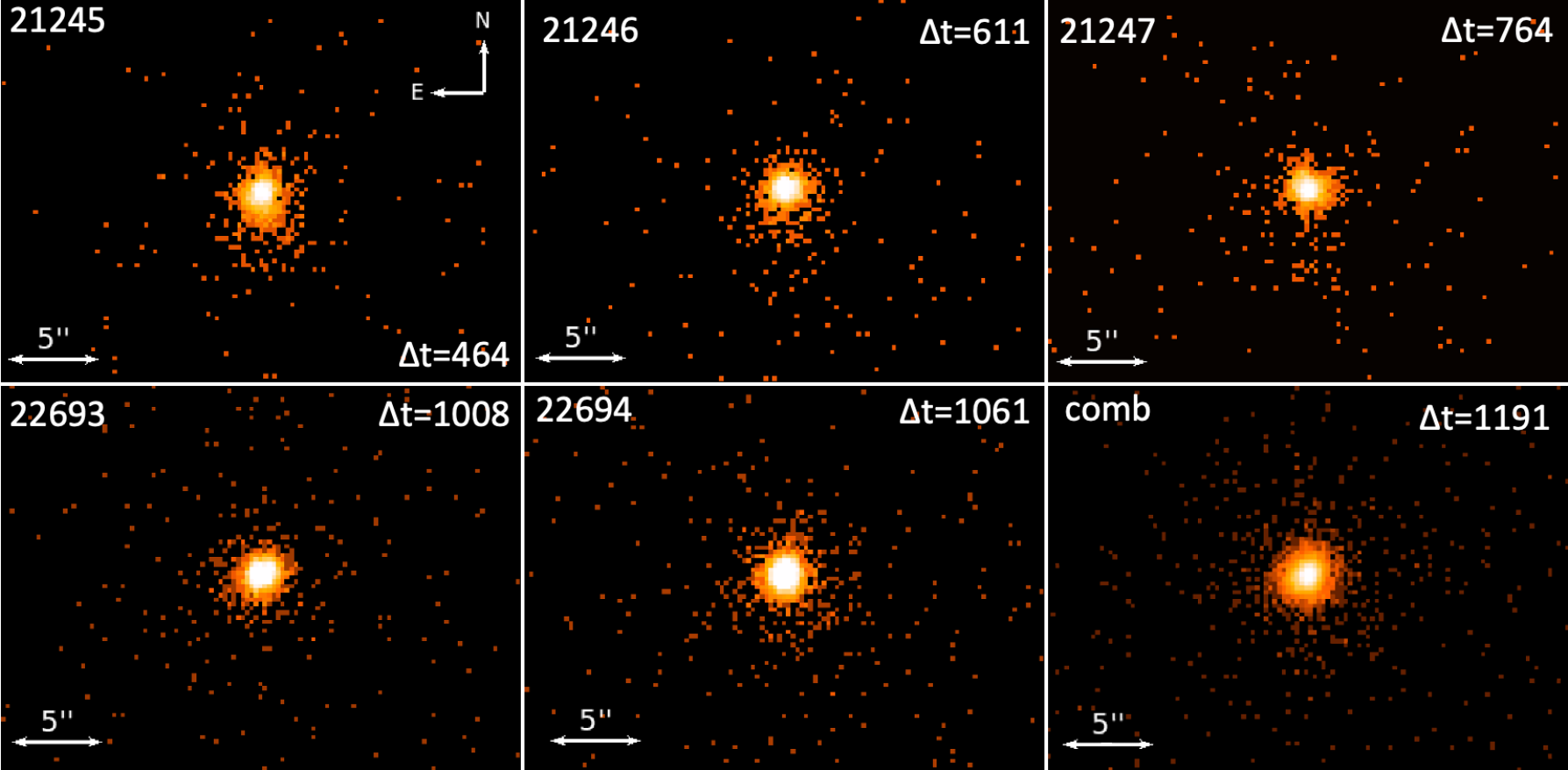}
\caption{Chandra ACIS 0.5-8 keV images from the 2017-2021 observing campaign with one half native pixel (0\farcs246) binning. The $\Delta t$ shows how many days after periastron (2017 September 22) the observation was taken. 
{
\label{new_obs_all}
}}
\end{figure*}

The series of six 
ACIS images of B1259 during the 2017--2021 binary cycle are shown in Figure \ref{new_obs_all}. It is apparent that a bright X-ray emitting clump detached from the binary, like 
those observed in previous binary cycles, was not seen in these observations. 
However, there are hints of extended asymmetric  emission around the binary, particularly in ObsIDs 21245, 21246, 21247, and 22694. To search for X-ray clumps buried in the wings of the point-spread function (PSF), we used the Lucy-Richardson deconvolution algorithm \citep{1972JOSA...62...55R,1974AJ.....79..745L}, which deconvolves the PSF from the intrinsic count distribution of the source. This was accomplished using the CIAO task {\tt arestore}, which requires an accurate PSF. The PSF model was created using the Chandra Ray Tracer (ChaRT) and was projected onto the ACIS-I detector plane using the MARX package \citep{2003ASPC..295..477C,2012SPIE.8443E..1AD}. MARX requires an input spectrum to model the energy dependence of the PSF based on the energy distribution of the source's counts. We fit the spectra of the binary with an absorbed power-law model and report the fitted model parameters in Table \ref{tab1}. 

We note that the binary spectrum from ObsID 21246, taken near apastron (see Figure \ref{obs_orbit}), is poorly fit for several reasons. The first is that there is an anomalously low spectral point below 1 keV, which causes  the absorbing column density to increase by a factor of 2-3. The $N_{\rm H}$ should not be changing by too much when the pulsar is moving through apastron, far from the high-mass companion and its decretion disk, making this increase in $N_{\rm H}$ unlikely to be real. This large $N_H$ value may be due to uncertainties in the calibration of the growing contamination layer on the ACIS optical blocking filters\footnote{See e.g., \url{https://cxc.harvard.edu/proposer/POG/html/chap6.html}}. Additionally, there are rather strong residuals in the spectrum around the absorption edges of iridium near 2-3 keV\footnote{see Figure 4.4 in the Proposer's Observatory Guide \url{https://cxc.cfa.harvard.edu/proposer/POG/pdf/MPOG.pdf}}, which also contribute to the poor fit. Simply ignoring energies below 1 keV and refitting the spectrum provides more consistent values and a chi-square decrease of $\Delta\chi^2=14$ for 1 less degree of freedom. This model gives $N_{\rm H}<3\times10^{21}$ cm$^{-2}$ (3$\sigma$ upper-limit), $\Gamma=1.50(9)$, $F_{\rm obs}=105(5)$, and $F_{\rm corr}=113(5)$ for $\chi^2/dof=101.1/69$.  When estimating the fluxes of the clump, we use the minimum observed absorbing column density (i.e., $N_{\rm H}=1.6\times10^{21}$ cm$^{-2}$), since the clump is far from the binary, thus should not experience a changing absorbing column density.

\begin{figure*}
\includegraphics[trim={0 0 0 0},width=18.0cm]{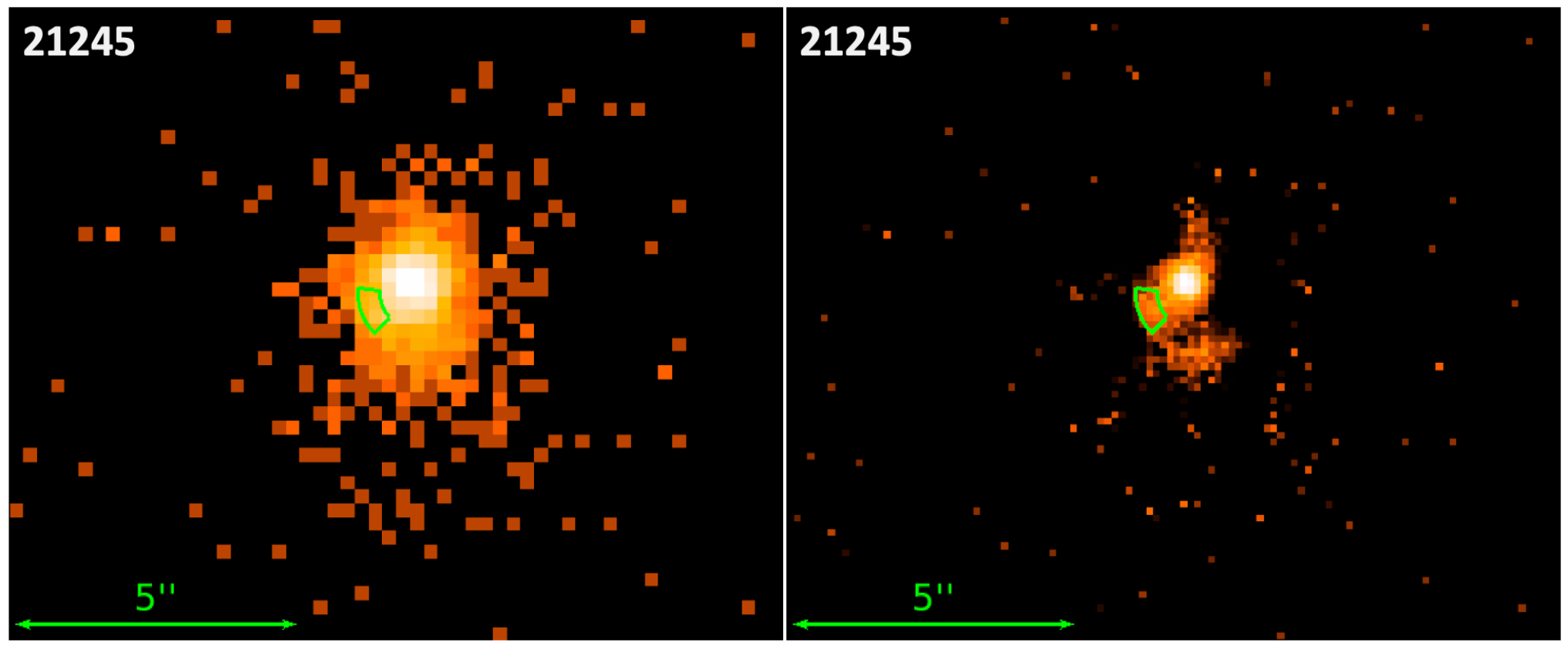}
\caption{{\sl Left:} ACIS 0.5-8 keV image of ObsID 21245 with native pixel binning. {\sl Right:} Deconvolved image of the same observation. The green region shows the image distortion caused by the Chandra mirror asymmetry. 
Extended emission can be clearly seen to the south of the binary.
{
\label{marx_21245}
}}
\end{figure*}

The strongest evidence for a clump 
was found in the deconvolved image of ObsID 21245 which occurred 463.5 days after periastron passage (see Figure \ref{marx_21245}). 
A hint of asymmetry is already seen in the original 
image, 
while the deconvolved image clearly shows a compact clump at a distance of about $1\farcs3$ south of the binary.
Unfortunately, the clump is too close to the binary to reliably estimate the number of net counts from the clump or to extract a spectrum. Furthermore, the deconvolution procedure does not conserve the number of counts in the deconvolved image, so it also cannot be used to estimate the number of counts. 

\begin{figure}
\includegraphics[trim={0 0 0 0},width=8.7cm]{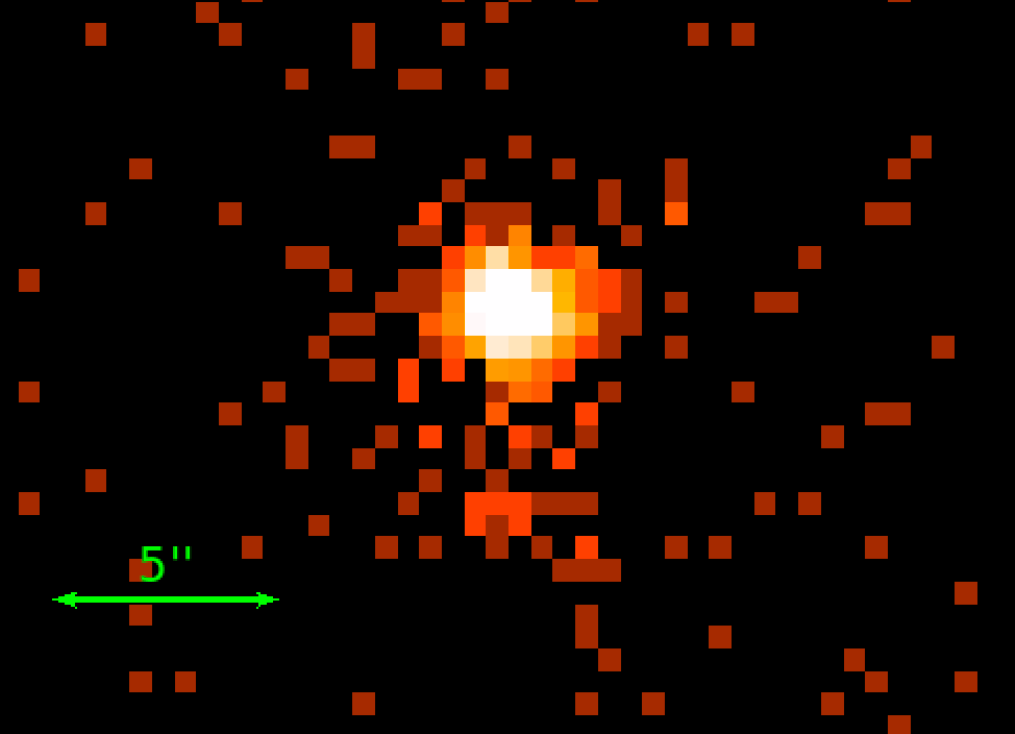}
\caption{ACIS 0.5--8 keV image of ObsID 21247 with native pixel binning. Faint extended emission can be seen to the south of the binary.  
{
\label{21247_bin}
}}
\end{figure}

There are also hints of faint extended emission that appear several arcseconds away from the binary in the images from ObsIDs 21246, 21247, and 22694. However, we do not see  strong evidence of extended emission near the binary in the deconvolved images from these ObsIDs. The lack of extended emission in the deconvolved images does not necessarily mean it is not there. The Lucy-Richardson deconvolution algorithm, used by CIAO's {\tt arestore} task, is known to amplify the background noise with each iteration, which can cause faint features (such as the faint clumps) to disappear in the deconvolved images (see, e.g., \citealt{2002PASP..114.1051S}). For these three observations, we used elliptical regions placed over the extended emission to estimate the numbers of net counts similar to those used in previous cycles (see Table \ref{tab1} and Figure 2 in \citealt{2019ApJ...882...74H}). We note that these numbers should be taken only as 
crude estimates as it is difficult to determine the shape of the extended emission, and the best extraction regions, due to its faintness. We estimate the flux from these detections by assuming an absorbed power-law model using the absorbing column density of the binary at apastron (since the extended emission is far from the companion wind/disk which causes variable absorption), and a photon index $\Gamma=1.4$ derived by \cite{2019ApJ...882...74H}. In ObsID 21247, where the clump is detected and clearly separated from the binary (see Figure \ref{21247_bin}), we find that 16 of the 21 photons have energies $>2$ keV, consistent with the previously measured spectrum.

As mentioned in the previous paragraph, it is difficult to determine the distance between the possible clump and binary in most observations 
because of the clump's faintness and proximity to the much brighter binary. However, in ObsIDs 21245 and 21247, the extended X-ray emission south of the binary 
is fairly concentrated, allowing 
a more accurate estimate of its distance from the binary. For ObsID 21245, we use the deconvolved image to estimate a distance of $1\farcs3$, whereas we use the original image to estimate a distance of  $4\farcs6$ between the binary and extended emission in ObsID 21247 (see Figures \ref{marx_21245} and \ref{21247_bin}). For the ObsIDs where there is no concentrated extended emission, we conservatively estimate upper limits by calculating the distance between the binary and the edge of any faint diffuse emission. We find upper limits of $3''$, $1\farcs8$, $3\farcs4$, and $4\farcs2$ for ObsIDs 21246, 22693, 22694, and the combined observation, respectively.

\begin{figure*}[ht!]
\includegraphics[trim={0 0 0 0},width=18.0cm]{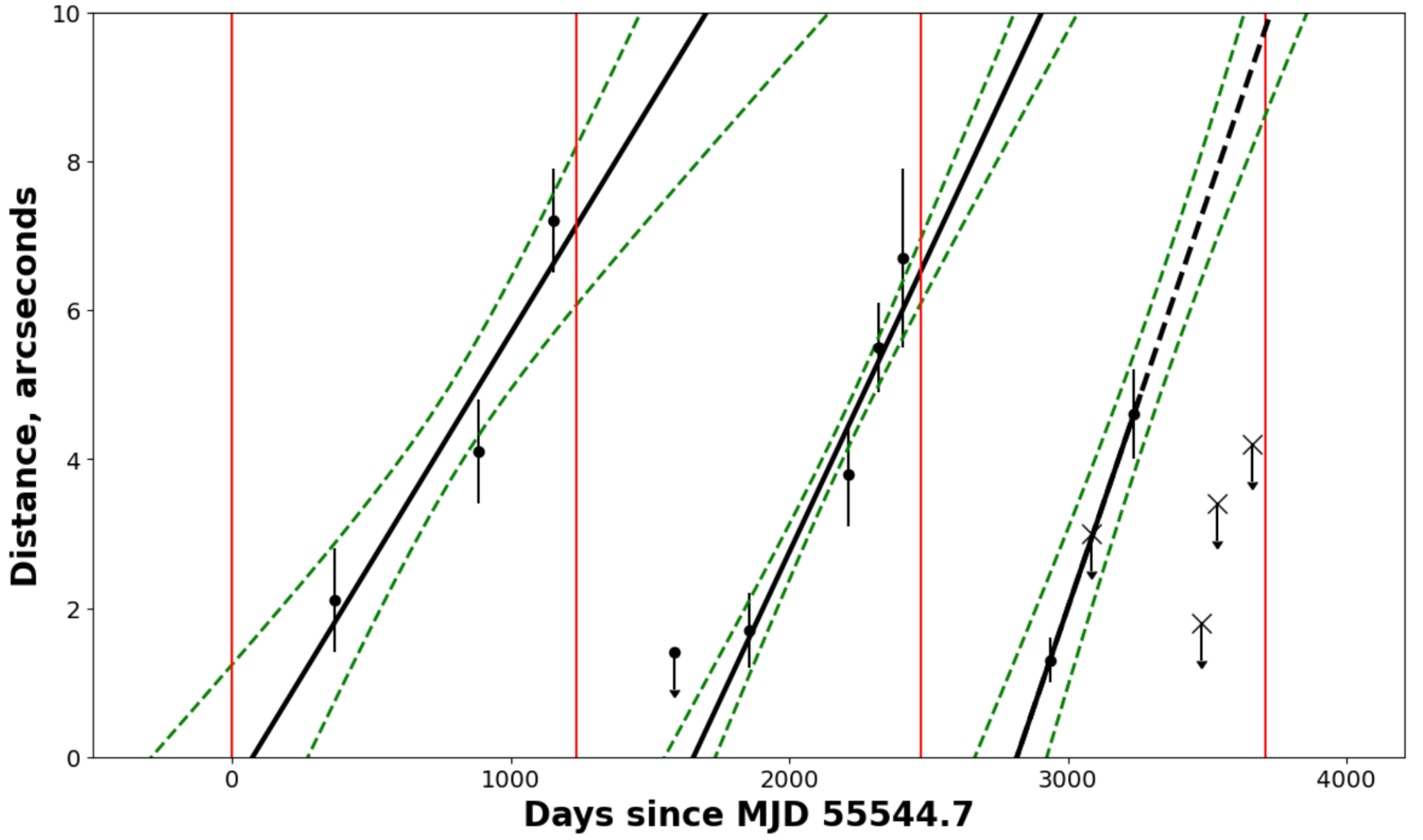}
\caption{The distance versus time for the X-ray emitting clump in 2011-2014, 2014-2017, and 2017-2021 binary cycles. The crosses mark the upper limit on the distance. The red lines show the times of periastron passage.}
\label{dist_v_time}
\end{figure*}

Similar to the previous binary cycles, we use the distance between the clump and the binary from ObsIDs 21245 and 21247 to constrain the velocity and launch time of the clump. However, 
we stress the important caveat that, unlike previous binary cycles where 
we clearly saw one 
moving and evolving 
clump, the same cannot be said for this cycle. Therefore, it is possible that the extended 
emission observed in ObsIDs 21245 and 21247 
did not come from the same clump,  
which would render these estimates inapplicable. If we assume 
that we saw the same clump in ObsIDs 21245 and 21247, which would also imply that the clump fades and brightens again since it is not detected in ObsID 21246, we can use their distances from the binary versus time to fit a linear model and obtain the projected velocity and launch time (see Figure \ref{dist_v_time}). Unfortunately, in this case, we only have two points (note that we do not use the upper limits) with uncertainties. We find a slope of $0\farcs011\pm0\farcs002$ day$^{-1}$, which corresponds to a projected velocity $v_\perp\approx(0.17\pm0.03)c\, d_{2.6}$.
The launch time for this model (i.e., the time when distance is equal to zero) is 345$\pm125$ days after periastron passage. A similar launch time was found by \cite{2019ApJ...882...74H} for the clump launched in the previous binary cycle. If the clump is instead launched at or near periastron, then \cite{2019ApJ...882...74H} show that the clump launched during that binary cycle may be accelerated. This assumption is reasonable, since it is difficult to understand how the material could be launched from the binary when the pulsar is far away from its companion. It is possible that the clump observed in this binary cycle is also accelerated, but 
we cannot exclude that we are seeing different 
clumps. We also lack enough firm measurements of the clump's position to constrain any acceleration.

This velocity is the highest we have observed from this system yet. However, we stress again that this could simply be due to the fact that the clumps detected in the two observations are not the same object. Assuming they are the same, this could imply that the clumps are less massive than those launched during previous binary cycles, and therefore, are traveling faster due to the constant ram pressure of the pulsar wind.

\section{Discussion}
\label{discuss}

The non-detection of a 
clearly detached, sufficiently bright clump 
during
the 2017--2021 binary cycle is rather surprising, given that such clumps have been observed in 
the previous  
two binary cycles \citep{2015ApJ...806..192P,2019ApJ...882...74H}. There is also evidence of another clump being launched during the current (2021--2024) binary cycle \citep{2023RNAAS...7...52H}, making 
2017--2021 the only binary cycle 
in which we do not definitively observe a fast-moving X-ray emitting clump. 

There are several potential scenarios that could 
explain the lack of such a clump that we discuss here. One scenario is that the properties of the 
decretion disk around the Be star lead to different disk mass/density 
each orbital period, which may, in turn, lead to different amounts of material being ejected from the binary.

The decretion disks of Be stars are known to be rather dynamic in a variety of ways. For instance, studies of the disks of Be stars have shown that they undergo growth and decay cycles over timescales of a few years, due to a varying mass injection rate from the star, which can influence the density and total mass of the material in the disk (see \citealt{2013A&ARv..21...69R} and references therein). One way these changes can be measured is by monitoring the variability in the equivalent width of the H$\alpha$ emission line, which is emitted by the decretion disk of the Be star. In fact, \cite{2016MNRAS.455.3674V} measured a decrease in the equivalent width of the H$\alpha$ emission line  between the spectra taken before the periastron passages in 2010 and 2014 and suggested that it may be due to a decrease in the disk mass. 
If this result is confirmed, it would indicate that the total mass of the decretion disk could change each orbit, leading to different amounts of material being ejected from the binary in a given orbital cycle. 

Another possible explanation for a varying amount of material being ejected from the binary is density waves in the Be star disk. The decretion disks of binary Be stars can become perturbed and grow spiral-like density waves that precess around the star on timescales of years \citep{2009A&A...504..929S,2009A&A...504..915C}. If such density waves exist in the decretion disk of LS 2883, then the pulsar's interaction with them may dictate how much material is ejected from the binary. In this case, the pulsar may have simply missed the high-density region of the wave  
and interacted with a low-density region,
ejecting less matter than in previous cycles. Further detailed studies of the Be disk of LS 2883, particularly when the pulsar is at apastron and the disk has recovered from the pulsar's passage, could help shed light on the orbit-to-orbit differences in the X-ray emitting ejecta, and may also help us to better understand the clump launching mechanism. Additional modeling of the interaction between the pulsar, pulsar wind, and decretion disk based on additional observations of the decretion disk can also help to shed light on the details of how the clump is launched and what factors dictate its evolution \citep[see, e.g.,][]{2011PASJ...63..893O}.

Another possible avenue to understanding the relationship between the properties of the X-ray emitting clumps and
the pulsar's interaction with the decretion disk is to study the 
multiwavelength behavior of the system 
around periastron (including 
the passage of the pulsar through the decretion disk). The lightcurve of B1259 typically shows two X-ray peaks attributed to the pulsar's passages through the Be star's decretion disk \citep{2014MNRAS.439..432C,2015MNRAS.454.1358C}. 
Fermi-LAT observations have shown that there are also GeV flares, seen only after periastron passages, with different delays and peak luminosities in different binary cycles 
(see, e.g., \citealt{2018ApJ...863...27J,2021Univ....7..472C,2021Univ....7..242C}). The X-ray emission during the 2017 periastron passage showed a delayed X-ray peak occurring $\sim5-10$ days after the pulsar crossed the decretion disk for the second time and a
GeV flare that was more delayed 
($\sim 60$ days after periastron in 2017 versus $\sim 30$ days in 2014 and 2010)
and had higher peak luminosities, some of which exceeded the spin-down luminosity of the pulsar \citep{2021Univ....7..472C}. It is still unclear exactly how (or even if) the differences in the X-ray and GeV light curves are correlated with the properties of the X-ray emitting clump (e.g., brightness, size). The X-ray and GeV light curves of the 2021 periastron passage showed some unique features, such as a third peak in the X-ray light curve and the most delayed, but intense, GeV emission observed yet from the binary \citep{2021ATel14612....1J,2021Univ....7..472C,2021Univ....7..242C}. Since another X-ray emitting clump has been detected after the 2021 periastron passage \citep{2023RNAAS...7...52H}, monitoring its size and brightness as it travels away from the binary 
will provide yet another opportunity to shed light on this fascinating phenomenon and system. 

\section{Conclusion}
\label{concc}

In conclusion, we observed PSR B1259-63 six times with Chandra after the 2017 periastron passage. In the deconvolved image of ObsID 21245, we find evidence that an X-ray emitting clump was launched from the binary. However, there are no additional firm detections of the clump in the later observations, but there is some faint extended emission surrounding the binary. ObsID 21247 shows a hint of extended emission which may be the X-ray emitting clump. Assuming this clump is the same clump that was observed in the first observation, we place constraints on the velocity ($v_\perp\approx(0.17\pm0.03)\,d_{2.6}\, c\, $) and launch time ($345\pm125$ days after periastron) of the clump. These values are consistent with those observed in previous binary cycles. A better understanding of the properties of the decretion disk in this system can help us to better model and understand the pulsar and pulsar wind's interaction with the disk. This can, in turn, lead to better models for the launching mechanism and evolution of these X-ray clumps.

\acknowledgements
 We thank the referee for carefully reading this paper and providing useful feedback. Support for this work was provided by the National Aeronautics and Space Administration through Chandra Awards GO9-20044X and GO1-22037X, and the Chandra ACIS Team contract SV4-74018, issued by the Chandra X-ray Center, which is operated by the Smithsonian Astrophysical Observatory for and on behalf of the National Aeronautics Space Administration under contract NAS8-03060. The Chandra Guaranteed Time Observations (GTO) data used here were selected by the ACIS Instrument Principal Investigator, Gordon P.\ Garmire, of the Huntingdon Institute for X-ray Astronomy, LLC, which is under contract to the Smithsonian Astrophysical Observatory, contract SV2-82024. J. H. acknowledges support from NASA under award number 80GSFC21M0002.  This paper uses the Chandra datasets, obtained by the Chandra X-ray Observatory, contained in~\dataset[DOI: https://doi.org/10.25574/cdc.159]{https://doi.org/10.25574/cdc.159}.

\facilities Chandra (ACIS; \citealt{2003SPIE.4851...28G})

\software Xspec \citep{1996ASPC..101...17A}, CIAO \citep{2006SPIE.6270E..60F}, SciPy \citep{2020NatMe..17..261V}, NumPy \citep{2020Natur.585..357H}

\end{document}